\documentclass[nofootinbib]{revtex4}%
\usepackage{amsmath}
\usepackage{graphicx}
\usepackage{dcolumn}
\usepackage{bm}
\usepackage{amssymb}
\usepackage{latexsym}

\def\be{\begin{equation}}
\def\ee{\end{equation}}
\def\bea{\begin{eqnarray}}
\def\eea{\end{eqnarray}}

\begin{document}

\title{Testing Oscillating Primordial Spectrum and Oscillating Dark Energy \\
with Astronomical Observations}
%\author{...}
\author{Jie Liu${}^a$}
\author{Hong Li${}^{a,b}$}
\author{Jun-Qing Xia${}^{c}$}
\author{Xinmin Zhang${}^{a,b}$}

\affiliation{${}^a$Institute of High Energy Physics, Chinese Academy
of Science, P.O. Box 918-4, Beijing 100049, P. R. China}

\affiliation{${}^b$Theoretical Physics Center for Science Facilities
(TPCSF), Chinese Academy of Science, P.R.China}

\affiliation{${}^c$Scuola Internazionale Superiore di Studi
Avanzati, Via Beirut 2-4, I-34014 Trieste, Italy}

\begin{abstract}
In this paper we revisit the issue of determining the oscillating
 primordial scalar power spectrum and oscillating
equation of state of dark energy from the astronomical observations.
By performing a global analysis with the Markov Chain Monte Carlo
method, we find that the current observations from five-year WMAP
and SDSS-LRG matter power spectrum, as well as the ``union"
supernovae sample, constrain the oscillating index of primordial
spectrum and oscillating equation of state of dark energy with the
amplitude less than $|n_{\rm amp}|<0.116$ and $|w_{\rm amp}|<0.232$
at $95\%$ confidence level, respectively. This result shows that the
oscillatory structures on the primordial scalar spectrum and the
equation of state of dark energy are still allowed by the current
data. Furthermore, we point out that these kinds of modulation
effects will be detectable (or gotten a stronger constraint) in the
near future astronomical observations, such as the PLANCK satellite,
LAMOST telescope and the currently ongoing supernovae projects SNLS.
\end{abstract}

\maketitle

\section{Introduction}

Recent advances in observational cosmology have revealed that our
universe has experienced at least two different stages of
accelerated expansion. One is the inflation in the very early
universe when its tiny patch was superluminally stretched to become
our observable Universe today. This can naturally explain why the
universe is flat, homogeneous and isotropic. Inflation is driven by
a potential energy of a scalar (or multi-scalar) called inflaton and
its quantum fluctuations turn out to be the primordial density
fluctuations which seed the observed large-scale structures (LSS)
and anisotropy of cosmic microwave background radiation (CMB). The
other one is accelerating expansion driven by dark energy (DE) which
dominates the energy density of the universe currently.
Understanding the nature of dark energy is among the biggest
problems in modern physics and has been studied actively in the
literature.

At present, the high quality observational data,
CMB\cite{wmap5,CBI,VSA}, LSS\cite{LSS} and type Ia supernovae (SN
Ia)\cite{Union} and so on, have provided the stringent constraints
on cosmological parameters. For example, current data can constrain
the primordial scalar power spectrum index $n_s$ and the energy
density of dark energy component $\Omega_{\rm de}$ to $1\%$ level
\cite{Xia:2008ex}. Besides the current observations, there are many
ongoing projects, such as PLANCK\cite{PLANCK}, LAMOST\cite{LAMOST}
and SNLS\cite{SNLS}. These projects will provide more accurate
measurements on CMB temperature anisotropies and polarization, LSS
matter power spectrum and the luminosity distance, which will be
helpful for studies of inflation and dark energy and determinations
of cosmological parameters.

Generally, the current data analysis bases on the simple
parameterizations of the primordial power spectrum and the equation
of state (EoS) of DE. However, we note that some inflation models
can generate the power spectrum with some modulated wiggles. This
picture can be realized by the inflaton field with a step function
of the potential \cite{star:1997,Adams:2001vc}or oscillating
potential\cite{Wang:2002hf}. The effects from the Trans-Planckian
initial condition can also lead to oscillations which do imprint
directly on the primordial scalar power
spectrum\cite{rb:trans,Danielsson:2002kx,Easther:2002mi}. The
bouncing model driven by the Quintom\cite{Feng:2004ad} matter can
also give rise to some wiggles in the primordial scale-invariant
power spectrum, because in this model the universe initially
experiences a contracting stage, after the contracting phase it
bounces to an inflationary phase, therefore the primordial
fluctuations in sub-hubble region would deviate from that generated
in Bunch-Davies vacuum\cite{cai:bounce}. The Quintom bounce provide
a solution to initial singularity problem, and in this scenario
there's no Trans-Planckian problem since we can choose the initial
condition via contracting phase. The featured primordial scalar
perturbation spectrum with local bumps are studied in
Ref.\cite{Kawasaki:2004pi}.

In some sense, the study for DE is similar to inflation, either in
model building or data fitting. The featured EoS of DE, especially
the oscillatory behavior EoS can provide us an unconventional
evolution of universe. Observationally, these kinds of modulated EoS
will leave clew on the hubble diagram or the matter power spectrum
as well as the temperature power spectrum of CMB, which give us some
hints to test such a scenario. The periodic oscillatory EoS can be
realized by the two scalar field Quintom matter\cite{Xiong:2008ic}.
And in Ref.\cite{Feng:2004ff}, a class of Quintom models with an
oscillating equation of state have been studied. In such a scenario,
the early inflation and the current acceleration of the universe can
be unified, the scale factor keeps increasing from one period to
another and leads naturally to a highly flat universe. The periodic
recurs will not lead to big crunch nor big rip and the coincidence
problem can be reconciled by the vicissitudinary repetition. Also,
the studied relevant to these kind of oscillating DE can be found in
\cite{Xia:2004rw,Xia:2006rr,Linder:2005dw,test}.

Given the above progress in the astronomical observations and
physical motivations, it can potentially lead us to study  issues
related to the featured structure of primordial spectrum and the EoS
of DE. In this paper, we aim to study the constraints on such kind
of featured parametrization of the primordial spectrum  as well as
the EoS of DE. We perform the global fitting by using the Markov
Chain Monte Carlo method with the current data from CMB, LSS and SN,
and also from the simulated future projects, such as PLANCK, 5-year
Supernovae Legacy survey (SNLS) and LAMOST.

The paper is organized as follows: In section II we introduce the
method of fitting and data we used in our analysis. The results and
discussions are given in section III. The last section is on the
summary.

%===================================================================

\section{method and data}

As mentioned before, there are many theoretical models which can
provide the non-trivial structure on the primordial power spectrum
and EoS of DE. In order to study the oscillating primordial scalar
power spectrum independent of the specific model, we parameterize
the power spectrum $P_{\chi}$ as: \be\label{paramtrize_inf}
\ln\mathcal{P}_{\chi}(k)= \ln
A_s(k)+\left[n_{s0}(k_{0})-1\right]\ln\left(\frac{k}{k_{0}}\right)-\frac{n_{\rm
amp}}{n_{\rm fre}}\cos\left[n_{\rm
fre}\ln\left(\frac{k}{k_{0}}\right)\right], \ee where, $A_s$ is the
amplitude of primordial power spectrum, $k_{0}$ is the scale pivot
and  is set as $0.05~\rm{Mpc}~ h^{-1}$. $n_{s0}$ characterizes the
tilt of spectrum while $n_{\rm amp}$ and $n_{\rm fre}$ denote the
contribution of featured oscillation. Comparing with the traditional
definition of spectral index, we get \be n_s =
\frac{d\ln\mathcal{P}_{\chi}(k)}{d\ln k} = n_{s0}+n_{\rm
amp}\sin\left[n_{\rm
fre}\ln\left(\frac{k}{k_0}\right)\right]\label{par_inf}, \ee and one
can find that for small $n_{\rm fre}$, this parametrization goes
back to the traditional form, i.e. the last term in the right hand
side of equation above gives rise to the running of the scalar
spectral index.

For the parametrization of DE, we take the form given in
\cite{xia:planck}: \be \label{par_eos}w(a)=w_0+w_{\rm
amp}\sin\left[w_{\rm fre}\ln(a)\right]. \ee This oscillating
behavior in the EoS can lead to the modulations on the Hubble
diagram or a recurrent universe which unifies the early inflation
and the current acceleration. In
Refs.\cite{Xia:2004rw,Xia:2006rr,Zhao:2006qg,Barenboim:2004kz} some
preliminary studies have been presented on this kind of DE model. We
noticed that there are some hints on the oscillating behavior for
example as shown in the Fig. 10 of the SNIa paper
\cite{Riess:2006fw} and see also paper in Ref. \cite{Wei:2008kq}.
Our sine function has the advantage of preserving the oscillating
feature of the DE EoS at high redshift measured by the CMB data. For
simplicity and focusing on the study at lower redshift, we set
$w_{\rm fre}$ to be $3\pi/2$ in order to allow the EoS to evolve
more than one period within the redshift range of $z=0$ to $z=2$,
where the SNIa data are most robust. For the $\Lambda$CDM model,
$w_0=-1$ and $w_{\rm amp}=0$.

We extend the publicly available MCMC package
CosmoMC\footnote{Available at:
http://cosmologist.info/cosmomc/.}\cite{CosmoMC}by including dark
energy perturbation\cite{Zhao:2005vj}, and  we assume the adiabatic
initial condition in our calculation. The most general parameter
space is: \be \label{parameter} {\bf P} \equiv (\omega_{b},
\omega_{c}, \Theta_{s}, \tau, w_0, w_{\rm amp}, n_{s0},n_{\rm
amp},n_{\rm fre},A_s), \ee where $\omega_{b}\equiv\Omega_{b}h^{2}$
and $\omega_{c}\equiv\Omega_{c}h^{2}$ with $\Omega_{b}$ and
$\Omega_{c}$ being the physical baryon and cold dark matter
densities relative to the critical density, $\Theta_{s}$ is the
ratio (multiplied by 100) of the sound horizon to the angular
diameter distance at decoupling, $\tau$ is the optical depth to
re-ionization, $w_0$ and $w_{\rm amp}$ are the EoS parameters of DE,
and $n_{s0},~n_{\rm amp},~n_{\rm fre},~A_s$ are the parameters
related to the primordial scalar power spectrum in Eq.
(\ref{paramtrize_inf}).

In our calculations, we take the total likelihood to be the
products of the separate likelihoods (${\bf \cal{L}}_i$) of CMB,
LSS and SNIa. Defining $\chi_{L,i}^2 \equiv -2 \log {\bf
\cal{L}}_i$, we then have \be\label{chi2} \chi^2_{L,total} =
\chi^2_{L,CMB} + \chi^2_{L,LSS} + \chi^2_{L,SNIa}~. \ee If the
likelihood function is Gaussian, $\chi^2_{L}$ coincides with the
usual definition of $\chi^2$ up to an additive constant
corresponding to the logarithm of the normalization factor of
${\cal L}$.

The data used for current constraints include the five-year WMAP
(WMAP5)\cite{wmap5} as well as some small-scale CMB measurements,
such as CBI \cite{CBI}, VSA\cite{VSA}, BOOMERanG\cite{BOOMERanG} and
the newly released ACBAR data\cite{ACBAR}. For the Large Scale
Structure information, we use the Sloan Digital Sky Survey (SDSS)
luminous red galaxy (LRG) sample\cite{Tegmark:2006az}. The
supernovae data we use are the recently released ``Union"
compilation of 307 samples \cite{Union}. In the calculation of the
likelihood from SNIa we marginalize over the relevant nuisance
parameter \cite{DiPietro:2002cz}. Furthermore, we make use of the
Hubble Space Telescope (HST) measurement of the Hubble parameter
$H_{0}\equiv 100$h~km~s$^{-1}$~Mpc$^{-1}$ \cite{HST} by multiplying
the likelihood by a Gaussian likelihood function centered around
$h=0.72$ and with a standard deviation $\sigma=0.08$.  We also
impose a weak Gaussian prior on the baryon density
$\Omega_{b}h^{2}=0.022\pm0.002$ (1 $\sigma$) from the Big Bang
Nucleosynthesis \cite{BBN}, and a cosmic age tophat prior as 10 Gyr
$< t_0 <$ 20 Gyr.

For the future data, we consider the measurements of LSS from
LAMOST\cite{LAMOST}, the CMB from PLANCK \cite{PLANCK} and the SN Ia
from 5-year SNLS\cite{SNLS}. For more information about the mock
data, we refer to \cite{xia:planck,li:lamost}.

For the simulation of LAMOST, we mainly consider the $3D$ matter
power spectrum of galaxies. The simulated error of the matter power
spectrum including the statistical errors due to sample variance and
shot noise are given by\cite{Feldman:1993ky} \be \label{eqn:dPK}
(\frac{\sigma_P}{P})^2 = 2\times \frac{(2 \pi)^3}{V}\times
\frac{1}{4 \pi k^2 \Delta k}\times (1+ \frac{1}{\bar{n}P})^2~, \ee
 where $V$ is the survey volume and $\bar{n}$ is the mean galaxy density.
In our simulations, we set the redshift of the LAMOST main sample to
be $z\sim 0.2$, and the survey area to be $15000$ $deg^2$. The total
number of galaxies within the survey volume is $10^7$\cite{LAMOST}.
The maximum k we consider is $k \sim 0.1$ $h$ Mpc$^{-1}$. For the
simulation with PLANCK, we follow the method given in our previous
paper \cite{xia:planck}. We mocked the CMB TT, EE and TE power
spectrum by assuming the certain fiducial cosmological model. For
the detailed techniques, we refer to our previous paper
\cite{xia:planck}. We have also simulated $500$ SN Ia according to
the forecast distribution of the SNLS \cite{Yeche:2005wn}. For the
error, we follow the Ref.\cite{Kim:2003mq} which takes the magnitude
dispersion $0.15$ and the systematic error $\sigma_{sys}=0.02\times
z/1.7$. The whole error for each data is given by:
\begin{equation}
\sigma_{maga}(z_i)=\sqrt{\sigma^2_{sys}(z_i)+\frac{0.15^2}{n_i}}~,\label{snap}
\end{equation}
where $n_i$ is the number of supernova of the $i'$th redshift bin.
We take $\Lambda$CDM model as the fiducial model in simulating the
data.

%\subsection{}

%\subsection{Global fitting program}

\section{Numerical Results}

\subsection{Oscillating Primordial Power Spectrum}
\begin{figure}
\begin{center}
\includegraphics[scale=0.4]{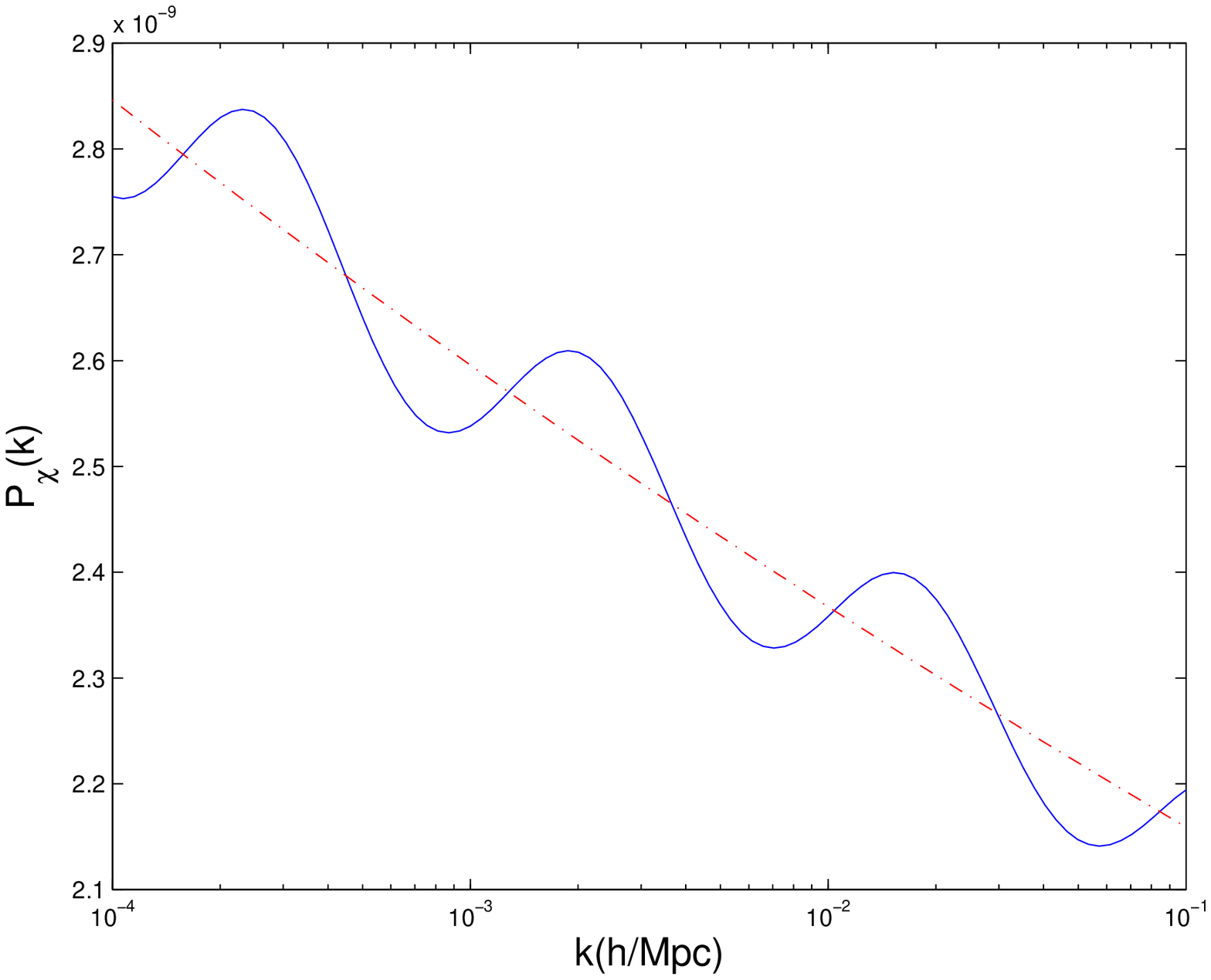}
\includegraphics[scale=0.4]{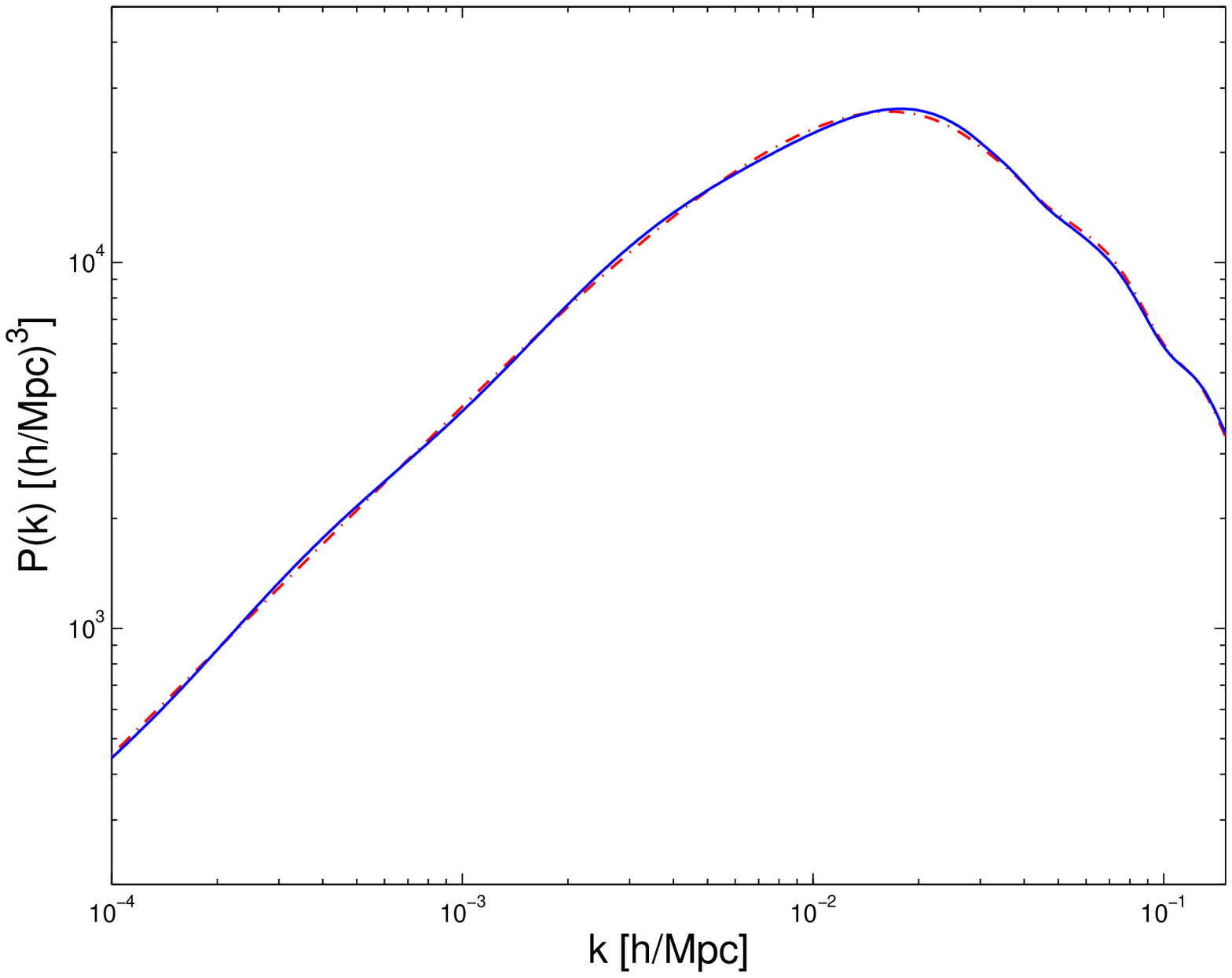}
\includegraphics[scale=0.4]{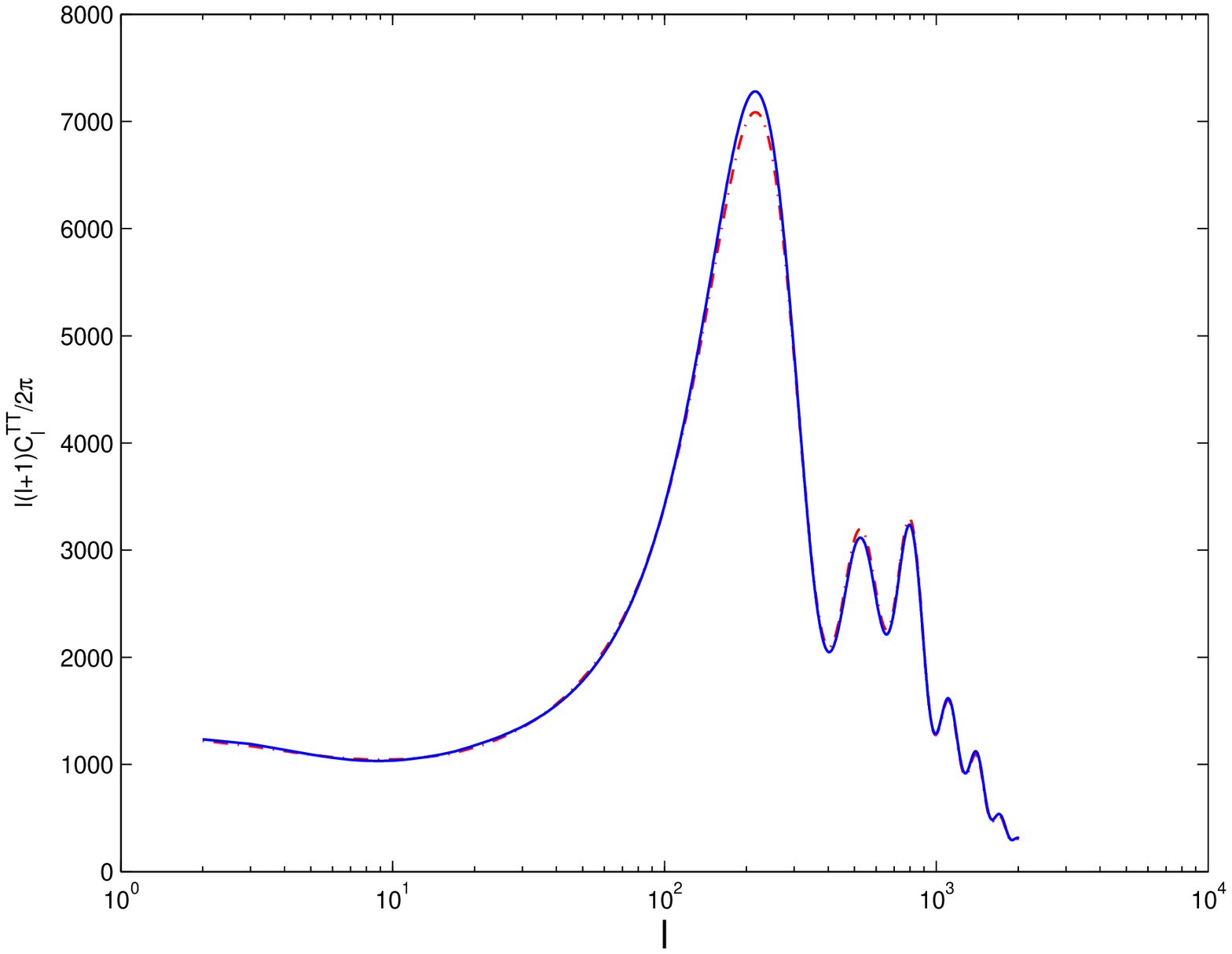}
\caption{Primordial power spectrum (top left), matter power spectrum
(top right) and CMB TT power spectrum(bottom) of the smooth
parametrization (i.e. parameterized by a power law with $n_s=0.96$
and $\alpha_s=0$) (red) and oscillating parametrization (blue). The
oscillating parameters are [$n_{s0}$, $n_{amp}$, $n_{fre}$]=[0.96,
0.10, 3.0] which are within $1\sigma$.}\label{comparison}
\end{center}
\end{figure}

The oscillating behavior of the primordial spectral index will
imprint on CMB and LSS via the following two equations:
\begin{align}
P_m(k)&=T^2(k)\mathcal{P}_{\chi}(k),\\
C_l&=\frac{4\pi}{2l+1}\int\frac{dk}{k}T^2_l(k)\mathcal{P}_{\chi}(k)~,
\end{align} where $T(k)$ and $T_l(k)$ are the transfer functions which denote
perturbation evolution of matter and photon from reheating era up to
now \cite{hu:trans,eisenstein:trans}. Based on the parametrization
in Eq. (\ref{paramtrize_inf}), in Fig. \ref{comparison}, we display
the matter power spectrum $P_m(k)$ and the TT power spectrum $C_l$.
The parameters are chosen to be [$n_{s0}$, $n_{amp}$,
$n_{fre}$]=[0.96, 0.10, 3.0], which can fit the current
observational data at $2\sigma$ C.L.. The blue solid lines are given
by the specific oscillatory power index, while the ordinary smooth
parametrization are given by the red dashed lines. One can find the
modulated oscillating structure in the whole linear regime of the
matter power spectrum, and this scenario as well as  the related
theoretical model has been considered in Ref.\cite{Wang:2002hf}.
Unlike Trans-Plankian mechanism or step-like models which just give
a local feature in the $k$ space, this kind of signature has effects
in the whole linear regime, however it might not be easy to
distinguish from baryon acoustic oscillation in small scales. From
Fig.\ref{paramtrize_inf} one can see the effects on CMB TT power
spectrum with the change of the amplitude of the  Doppler peaks. In
short, these distinct hints on the observations will feed back
constraints on the parameters.

In Table \ref{result}, we give the main results of our numerical
calculation. We present the $1, 2\sigma$ constraints from the
current data which include WMAP5, SDSS-LRG matter power spectrum as
well as the ``union" SN Ia data. We also provide the constraints
from the simulated PLANCK TT, TE and EE power spectrum, the LAMOST
matter power spectrum and the 5-year SNLS data. For the future
constraints, we mainly give the standard deviation.

\begin{table}[htbp]
\begin{tabular}{|c|c|c|c|}
  \hline
  ~~~&Current Constraints& Future Constraints\\
  \hline
  $n_{s0}$ & $0.974\pm0.016\pm0.033$  & $0.003$  \\
  \hline
  $n_{\rm amp}$ & $0.0\pm0.066\pm0.116$& $0.028$  \\
  \hline
  $n_{\rm fre}$ & $<6.256$ & $<1.651$  \\
  \hline
  $w_0$ & $-0.958_{-0.098-0.230}^{+0.098+0.161}$ & $0.040$  \\
  \hline
  $w_{\rm amp}$ & $0.030^{+0.124+0.232}_{-0.130-0.276}$ & $0.140$ \\
  \hline

\end{tabular}

\centering \caption{Constraints on the cosmological parameters from
the current and future observations. Here we show the mean values
and $1,2\sigma$ error bars. For the future measurements, we give the
standard deviation of these parameters. For some parameters that are
only weakly constrained we quote the $95\%$ upper
limit.}\label{result}
\end{table}

\begin{figure}
\begin{center}
\includegraphics[scale=0.4]{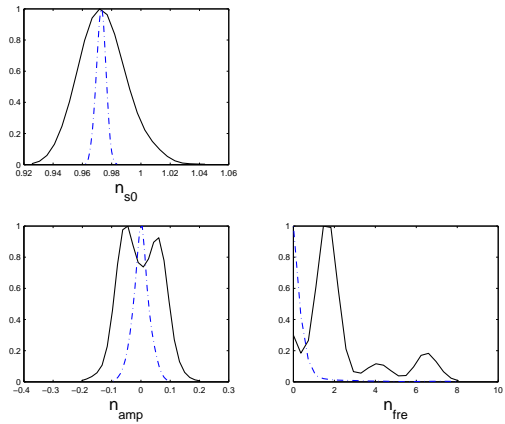}
\caption{One-dimensional constraints on the $n_{amp}$(top panel) and
$n_{fre}$(bottom panel). The black solid lines are given by the
current data, while the blue dashdot lines are given by the
simulated data:SNLS$+$PLANCK$+$LAMOST.}\label{ospp}
\end{center}
\end{figure}

In Fig \ref{ospp}, we give the 1-D probability distribution of
$n_{s0}$, $n_{\rm amp}$ and $n_{\rm fre}$. The black solid line is
the constraints from the current observational data. We get
$n_{s0}=0.974\pm0.016(1\sigma)$, and also the constraints on the
amplitude of the power index $|n_{\rm amp}|<0.116$ at $95\%$
confidence level. The scale invariant power spectrum is within
$2\sigma$ C.L., but the constraints on the frequency of power index
is still very weak, namely, the $2\sigma$ upper limit is
$n_{fre}<6.256$. Therefore, the current data still allow the
oscillating structures on the primordial scalar spectrum.

Since the present data do not give very stringent constraints on the
parameters, it is worthwhile discussing whether future data could
determine these parameters conclusively. From the Table
\ref{result}, one can see that the constraints get tightened
significantly. The standard deviation of $n_{s0}, n_{amp}$ are
$0.003$ and $0.028$ respectively, while the $2\sigma$ upper limit of
$n_{fre}$ is $1.651$. Since mocked data are generated from the
fiducial model with the standard power-law primordial spectrum, the
best fit values of $n_{amp}$ and $n_{fre}$ are expected to be around
zero. With this simulation, we find the scale invariant power
spectrum can be tested at $8\sigma$ C.L., and the error of the
oscillating amplitude can be shrunk by a factor of $2.35$.

\subsection{Oscillating dark energy}

The main contribution of dark energy on CMB is on the geometrical
angular diameter distance to the last scattering surface. The
oscillating EoS can also modulate the TT power spectrum as well as
the matter power spectrum, since oscillating EOS of DE can leave
somewhat similar imprints as oscillating primordial spectrum. For
the evolution of EoS given in Eq.(\ref{par_eos}), when $w_0$
deviates significantly from $-1$ and $w$ is matter-like, the
contributions to large scale CMB are significant due to the
Integrated Sachs-Wolfe (ISW) effects.

Firstly, with current data, we get $w_0=-0.958\pm0.098$,
$w_{amp}=0.030^{+0.124}_{-0.130}$ at $68\%$ confidence level,
however, $\Lambda CDM$ remains a good fit.

\begin{figure}
\begin{center}
\includegraphics[scale=0.4]{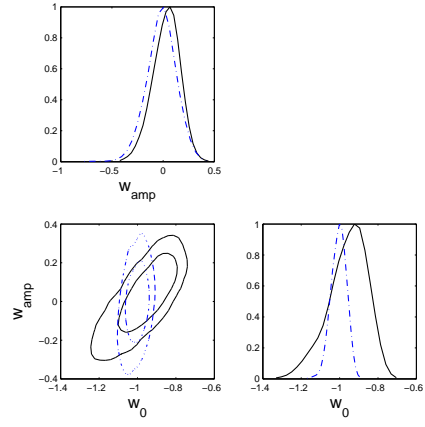}
\caption{One-dimensional constraints on the $w_0$(bottom right
panel) and $w_{\rm amp}$(top panel). The black solid lines are given
by current data, the blue dashdot lines are given by mocked
data:SNLS$+$PLANCK$+$LAMOST.}\label{osqt1d}
\end{center}
\end{figure}

\begin{figure}
\begin{center}
\includegraphics[scale=0.4]{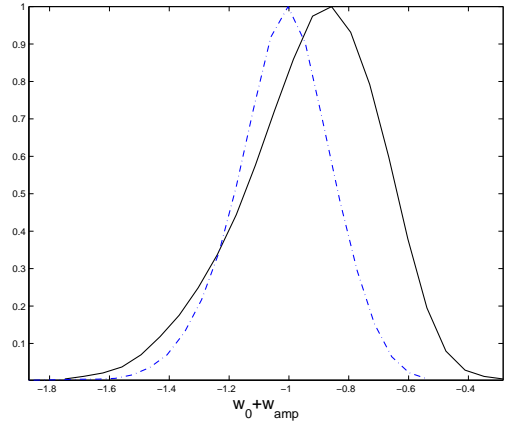}
\caption{One-dimensional constraints on the $w_0+w_{\rm amp}$.
Different colored line represents different datasets used as
before.}\label{w0pwa}
\end{center}
\end{figure}

In Fig. \ref{osqt1d} we display the two dimensional cross-correlated
constraints and one dimensional probability distribution of $w_0$
and $w_{amp}$.  The black solid lines are from the current data
sets. For the future simulated data, the $w_0$ will be constrained
tightly,  however $w_{amp}$ gets broaden  a little bit, which is due
to the degeneracy between the $w_0$ and the $w_{amp}$.
Quantitatively, the future simulated data shrinks the error bar of
$w_0$ by a factor of $2.4$ as delineated by the blue dashdot lines.
And interestingly, it also rotates the contour in the
$w_0$-$w_{amp}$ plane. To understand this phenomenon we notice that
physics which is interesting to us is the possible deviation of the
$w$ from $w=-1$. In our parametrization, $w_0+w_a$ gives the largest
value for this deviation. In Fig. \ref{w0pwa}, we plot the 1-D
probability distribution of $w_0+w_a$. From this figure one can see
indeed the future data give a tighter constraint.

\begin{figure}
\begin{center}
\includegraphics[scale=0.4]{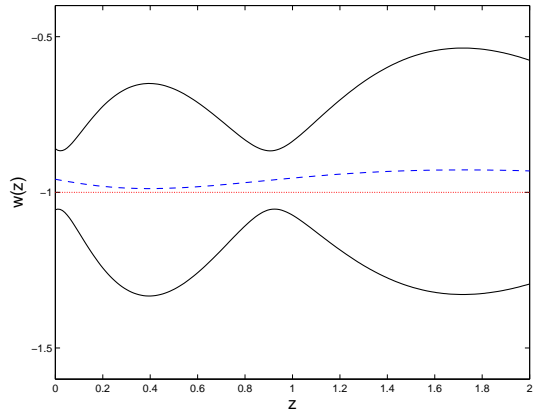}
\caption{Constraints on $w(a)=w_0+w_{\rm
amp}\sin[\frac{3\pi}{2}\ln(a)]$ from the current observations.
Median value (central red dotted line), 68\% (area between black
solid lines) intervals are shown. The red dotted line is the
cosmological constant boundary.}\label{err}
\end{center}
\end{figure}

Finally, in figure \ref{err} we present the one dimensional
constraint on the evolution of $w(z)$ from the current data. One can
see from this figure $\Lambda CDM$ remains a good fit, however the
dynamical models with oscillated feature are not excluded.
%the This behavior
%can be found more obviously from the best fit model. However,
%current data can not distinguish different dark energy models
%decisively, namely, the variance of $w_0$ and $w_{amp}$ are too
%large to distinguish dynamical dark energy models from the $\Lambda
%CDM$ model.

\section{Summary}
In this paper we have studied the signature of the modulated
primordial scalar power spectrum and the oscillating EoS of DE in
the observations, such as CMB, LSS and SN Ia. Based on the
parameterizations of $n_s(k)$ and $w(z)$ in Eq.(\ref{par_inf}) and
Eq. (\ref{par_eos}), we firstly present the constraints with the
current observations. Furthermore, we consider the constraints from
the future astronomy surveys, for example, the future CMB project
PLANCK, the LAMOST telescope and $500$ SN from the coming $5$ year
SNLS.

Our results show that the current data have put constraints on the
featured primordial scalar power spectrum scenario, however, the
limits are rather weak. From global data analysis, we get
$n_{s0}=0.974\pm0.016$ which indicates a red power spectrum
consistent with our previous analysis. Within $2\sigma$ C.L., with
the current observations we have $|n_{\rm amp}|<0.116$ and
$n_{fre}<6.256$, and for the future observations, the error bar can
be shrunk significantly.

On the other hand, for the constraints on DE, we find the
oscillating EoS of DE can fit the current data well, as well as the
$\Lambda CDM$ model. The amplitude of the oscillation of EoS are
limited to be $|w_{\rm amp}|<0.232$ at $95\%$ confidence level. The
future data will give a stronger constraint, especially on the $w_0$
parameter.

\acknowledgments

Our MCMC chains were finished in the Shuguang 4000A of the Shanghai
Supercomputer Center (SSC). We thank  Robert Brandenberger, Yi-Fu
Cai and Tao-Tao Qiu for helpful discussions. This work is supported
in part by National Science Foundation of China under Grant Nos.
10803001, 10533010 and 10675136, and the 973 program No.
2007CB815401, and by the Chinese Academy of Science under Grant No.
KJCX3-SYW-N2.

%%%%%%%%%%%%%%%%%%%%%%%%%%%% Referrence %%%%%%%%%%%%%%%%%%%%%%%%%%%%%

\end{document}